\documentclass[12pt,preprint]{emulateapj}

\begin{document}

\shorttitle{X-rays from Millisecond Pulsars in M28}
\shortauthors{Bogdanov et al.}

\title{\textit{Chandra} X-ray Observations of 12 Millisecond Pulsars \\ in the Globular Cluster M28}

\author{Slavko Bogdanov\altaffilmark{1,2}, Maureen van den
  Berg\altaffilmark{3}, Mathieu Servillat\altaffilmark{3}, Craig
  O.~Heinke\altaffilmark{4}, \\ Jonathan E.~Grindlay\altaffilmark{3},
  Ingrid H.~Stairs\altaffilmark{5}, Scott M.~Ransom\altaffilmark{6},
  Paulo C.~C.~Freire\altaffilmark{7}, \\ Steve B\'egin\altaffilmark{5,8},
  Werner Becker\altaffilmark{9}}

\altaffiltext{1}{Department of Physics, McGill University, 3600 University Street, Montreal, QC H3A 2T8, Canada; bogdanov@physics.mcgill.ca}

\altaffiltext{2}{Canadian Institute for Advanced Research Junior Fellow}

\altaffiltext{3}{Harvard-Smithsonian Center for Astrophysics, 60 Garden St., Cambridge, MA 02138, USA}

\altaffiltext{4}{Department of Physics, University of Alberta, 11322 89 Avenue, Edmonton, AB T6G 2G7, Canada}

\altaffiltext{5}{University of British Columbia, 6224 Agricultural Road Vancouver, BC V6T 1Z1, Canada}

\altaffiltext{6}{National Radio Astronomy Observatory, 520 Edgemont Road,
Charlottesville, VA 22901, USA}

\altaffiltext{7}{Max-Planck-Institut f\"ur Radio Astronomie, Auf dem H\"ugel 69, D-53121 Bonn, Germany}

\altaffiltext{8}{D\'epartement de physique, de g\'enie physique et d'optique, Universit\'e Laval, Quebec, QC G1K 7P4, Canada}

\altaffiltext{9}{Max-Planck-Institut f\"ur Extraterrestrische Physik, D-85740 Garching bei M\"unchen, Germany}

\begin{abstract}
  We present a \textit{Chandra X-ray Observatory} investigation of the
  millisecond pulsars (MSPs) in the globular cluster M28 (NGC
  6626). In what is one of the deepest X-ray observations of a
  globular cluster, we firmly detect seven and possibly detect two of
  the twelve known M28 pulsars.  With the exception of PSRs B1821--24
  and J1824--2452H, the detected pulsars have relatively soft spectra,
  with X-ray luminosities $10^{30-31}$ ergs s$^{-1}$ (0.3--8 keV),
  similar to most ``recycled'' pulsars in 47 Tucanae and the field of
  the Galaxy, implying thermal emission from the pulsar magnetic polar
  caps.  We present the most detailed X-ray spectrum to date of the
  energetic PSR B1821--24. It is well described by a purely
  non-thermal spectrum with spectral photon index $\Gamma=1.23$ and
  luminosity $1.4\times10^{33}\Theta(D/5.5~{\rm kpc})^2$ ergs s$^{-1}$
  (0.3--8 keV), where $\Theta$ is the fraction of the sky covered by
  the X-ray emission beam(s). We find no evidence for the previously
  reported line emission feature around 3.3 keV, most likely as a
  consequence of improvements in instrument calibration. The X-ray
  spectrum and pulse profile of PSR B1821--24 suggest that the bulk of
  unpulsed emission from this pulsar is not of thermal origin, and is
  likely due to low-level non-thermal magnetospheric radiation,
  an unresolved pulsar wind nebula, and/or small-angle scattering of
  the pulsed X-rays by interstellar dust grains.  The peculiar binary
  PSR J1824-2452H shows a relatively hard X-ray spectrum and possible
  variability at the binary period, indicative of an intrabinary shock
  formed by interaction between the relativistic pulsar wind and
  matter from its non-degenerate companion star.

\end{abstract}


\section{INTRODUCTION}

Rotation-powered millisecond pulsars (MSPs) are faint X-ray sources,
with typical luminosities of $L_X\sim10^{30-31}$ ergs s$^{-1}$ and up
to $\sim10^{33}$ ergs s$^{-1}$. Due to the long exposures required to
study MSPs in detail, globular clusters provide a particularly
efficient way to observe them in large numbers. This has only become
feasible in the past decade, owing to the sub-arcsecond angular
resolution of the \textit{Chandra X-Ray Observatory} that is essential
for individually resolving the numerous X-ray sources in the dense
cores of clusters.  Thus far, a total of $\sim$30 X-ray counterparts
of MSPs have been identified with \textit{Chandra} in globular
clusters, including 47 Tuc \citep{Grind02,Bog06}, NGC 6397
\citep[][]{Grind02,Bog10}, M28 \citep{Beck03}, M4 \citep{Bass04a}, NGC
6752 \citep{DAmico02}, M71 \citep{Els08}, and Terzan 5
\citep{Hein06}. This appreciable sample has already provided important
insight into the general X-ray properties of the MSP population in the
Galaxy.

The globular cluster Messier 28 (NGC 6626) is at a distance
of $D\approx5.5$ kpc.  The reddening towards M28 $E(B-V)=0.43$
\citep{Harris96} implies a hydrogen column density of $N_{\rm H}
\approx 2.4\times10^{21}$ cm$^{-2}$. For over two decades, M28 was
known to host a single millisecond pulsar (MSP), PSR B1821--24
\citep{Lyne87}.  Recently, the Green Bank Telescope detected
11 additional radio pulsars in M28 \citep{Beg06} making it the cluster
with the third largest number of known pulsars to date after Terzan 5
\citep[at least 33, see][]{Ran05} and 47 Tucanae \citep[23, see][and
references therein]{Cam00,Freire03}\footnote{For an up-to-date list of
  all known globular cluster MSPs see
  \url{http://www.naic.edu/\~pfreire/GCpsr.html}.}. 
These new pulsar discoveries make M28 a well-suited globular cluster
for effective X-ray studies of a sizable population of MSPs.

\citet{Beck03} have previously reported on a set of $\sim$40.9 ks
\textit{Chandra X-ray Observatory} ACIS-S observations of M28 (ObsIDs
2683, 2684, 2685) which probed down to a luminosity of
$L_X\sim6\times10^{30}$ ergs s$^{-1}$ (assuming a source detection
threshold of $\sim$10 counts) and detected 46 sources, only 12 of
which are within one core radius (0.24$'$) and 26 within the half-mass
radius (1.56$'$). Due to the relatively short exposure, just six
sources had sufficient counts for a spectral fitting analysis, which
included the original globular cluster MSP, PSR B1821--24.
\citet{Rut04} have used the superb timing capabilities of the HRC-S
instrument on board \textit{Chandra} to study the X-ray pulsations
from B1821--24. More recently, this MSP was again targeted with HRC-S
to calibrate the absolute timing precision of this detector
(unpublished data).

%
%
\begin{figure*}[t]
\begin{center}
\includegraphics[width=0.1237\textwidth]{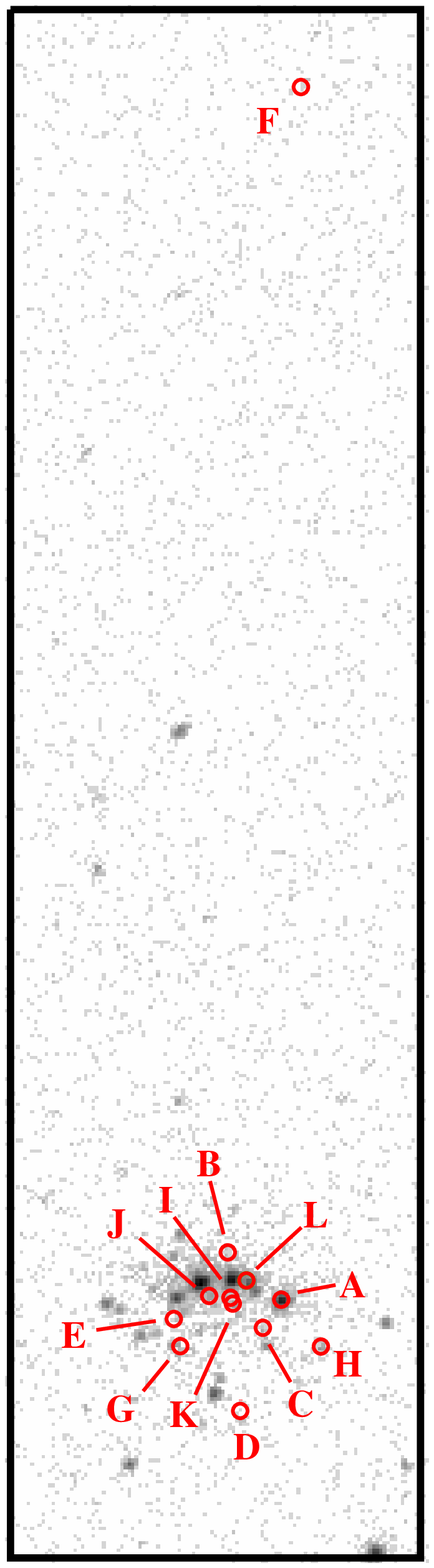}
\includegraphics[width=0.45\textwidth]{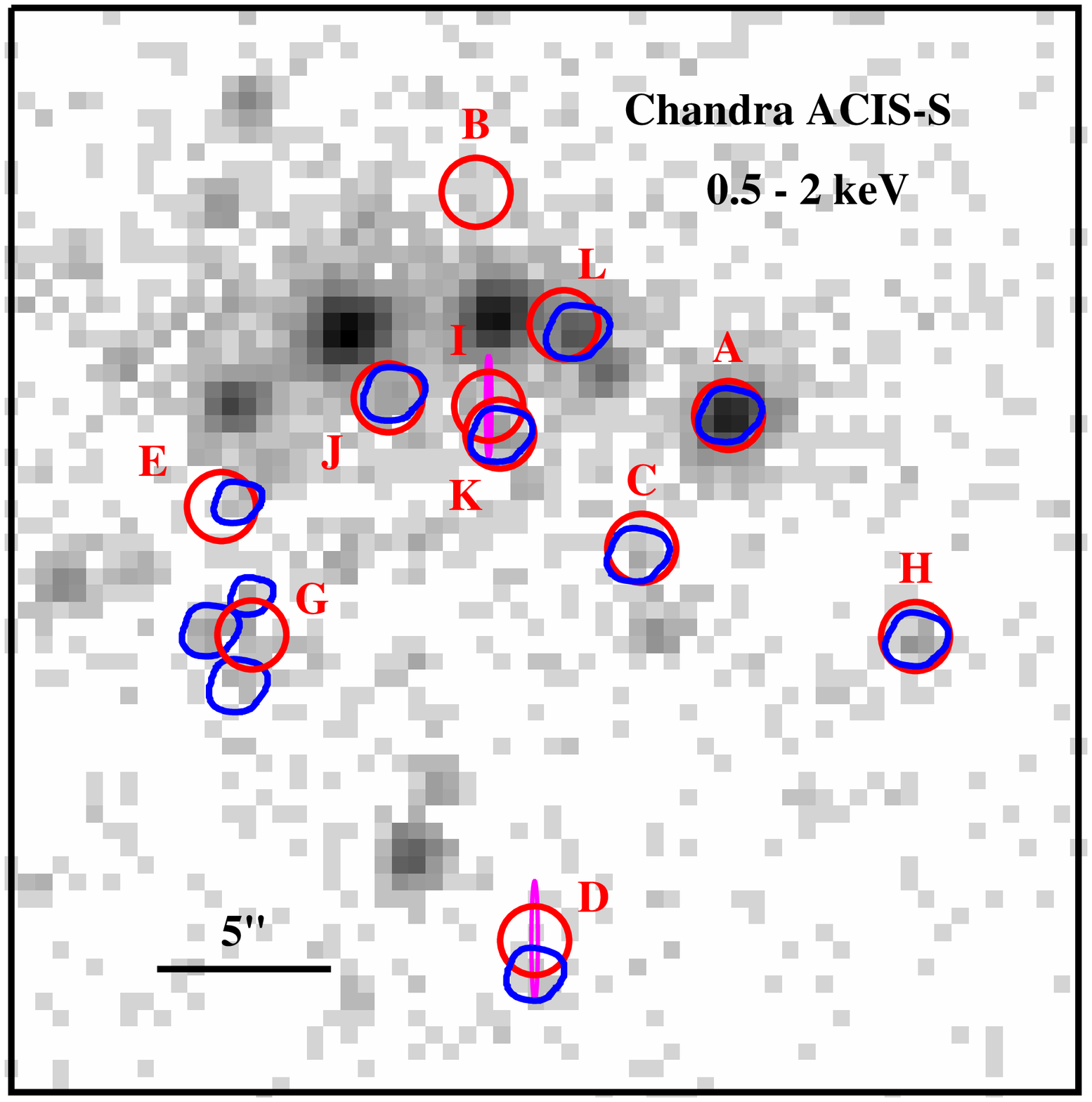}
\end{center}
\caption{(\textit{Left}) \textit{Chandra} ACIS-S coadded 0.5--2 keV
  image of the globular cluster M28 with the positions of the 12 known
  radio pulsars in the cluster labeled. (\textit{Right}) The inner
  core of M28, with 1$\arcsec$ radius red circles centered on the
  radio pulsar timing positions and blue polygons representing the
  extraction regions of the nearest X-ray source to each pulsar, with
  the exception of pulsar G for which the nearest three sources are
  shown. The magenta ellipses show the TEMPO radio timing position
  uncertaint for pulsars D and I, multiplied by a factor of 2 and 5,
  respectively (see text for details). For all other pulsars, the
  radio position uncertainties are much smaller
  ($\lesssim$0.1$\arcsec$). The image greyscale corresponds to number
  of counts increasing logarithmically from 0 (white) to 2553
  (black). North is up and east is to the left.}
\end{figure*}

In this paper, we present new deep \textit{Chandra X-ray Observatory}
ACIS-S observations of M28 and its 12 known pulsars. This
investigation sheds further light onto the general X-ray properties of
the Galactic population of MSPs. A comprehensive analysis of the other
X-ray source classes in M28 will be presented elsewhere (M.~van den
Berg, et al., in preparation). The present work is outlined as
follows. In \S 2, we describe the observations and data reduction and
analysis procedures. In \S 3 we present the imaging analysis and
detection methods of the X-ray counterparts to the MSPs.  In \S4, we
conduct a spectroscopic and timing study of PSR B1821--24 (M28 A), while in
\S5 we focus on the peculiar binary PSR J1824--2452H (M28 H). In \S6 we
discuss the remaining pulsars in the cluster. We offer conclusions in
\S7.

\begin{deluxetable}{lccc}
\tabletypesize{\small} 
\tablecolumns{5} 
\tablewidth{0pc}
\tablecaption{Chandra Observations of M28}
\tablehead{ \colhead{Telescope/} &\colhead{Epoch of} & \colhead{Observation} & \colhead{Exposure} \\
\colhead{Instrument} & \colhead{Observation} & \colhead{ID} &
\colhead{Time (ks)}}
\startdata
Chandra/ACIS-S & 2002 Jul 4  & 2684 & 12.9 \\
Chandra/ACIS-S & 2002 Aug 4  & 2685 & 13.7 \\
Chandra/ACIS-S & 2002 Sep 9  & 2683 & 14.3 \\
Chandra/HRC-S  & 2002 Nov 8  & 2797 & 49.4 \\
Chandra/HRC-S  & 2006 May 27 & 6769 & 41.1 \\
Chandra/ACIS-S & 2008 Aug 7  & 9132 & 144.1 \\
Chandra/ACIS-S & 2008 Aug 10 & 9133 & 55.2
\enddata
\end{deluxetable}

\section{OBSERVATION AND DATA REDUCTION}

The new \textit{Chandra} dataset presented herein was acquired during
two separate observations in Cycle 9, on 2008 August 7 (ObsID 9132)
and 2008 August 10 (ObsID 9133) for 144.1 and 55.2 ks, respectively.
For both exposures, the optical center of the cluster at $\alpha=
18^{\rm h}24^{\rm m}32.89^{\rm s}$ and
$\delta=-24^{\circ}52\arcmin11.4\arcsec$ \citep[][]{Shawl86} was
positioned 1$\arcmin$ off-axis from the nominal aim point of the
back-illuminated ACIS-S3 CCD, configured in VFAINT telemetry mode. We
also include the three Cycle 3 ACIS-S observations of 14.3, 12.9, and
13.7 ks (ObsIDs 2683, 2684, and 2685, respectively) from
\citet{Beck03}, also taken in VFAINT mode. In the analysis of PSR
B1821--24 (see \S4) we make use of the two HRC-S exposures of 49.4 and
41.1 ks (ObsIDs 2797 and 6769, respectively) in the special SI mode,
which permits a timing resolution of $\sim$16 $\mu$s. All observations
used herein are summarized in Table 1.

The data re-processing, reduction, and analysis were performed using
CIAO\footnote{Chandra Interactive Analysis of Observations, available
  at \url{http://cxc.harvard.edu/ciao/}} 4.1.2.  Starting from the
level 1 data products of the ACIS-S observations, we first removed
pixel randomization from the standard pipeline processing to aid in
source disentanglement in the crowded cluster core. In addition, for
the purposes of faint source detection we applied the background
cleaning algorithm specific to the VFAINT telemetry mode. However,
this procedure tends to reject real source counts for relatively
bright sources. Consequently, the background-cleaned data were not
used for the spectral analyses discussed in \S 4 and 5. Before
coadding the five ACIS-S images, the individual aspect solutions were
reprojected relative to the longest observation (ObsID 9132) using the
brightest sources in the cluster to correct for any differences in the
absolute astrometry between the individual observations. The summed
ACIS-S data results in a 237.1 ks effective exposure time, one of the
deepest X-ray observations of a globular cluster, with a limiting
sensitivity of $L_X\sim1\times10^{30}$ ergs s$^{-1}$ for a 1 keV
source at the distance of $D=5.5$ kpc and hydrogen column density
$N_{\rm H}=2.4\times10^{21} $ cm$^{-2}$.

For the spectroscopy and variability analyses of the X-ray
counterparts to the MSPs, we extracted the emission from polygonal
(roughly circular) regions that enclose $\sim$90\% of the total source
energy at 1.5 keV.  To permit spectral fitting in
XSPEC\footnote{Available at
  \url{http://heasarc.nasa.gov/docs/xanadu/xspec/index.html}.}, the
extracted source counts in the 0.3--8 keV range were grouped in energy
bins so as to ensure at least 15 counts per bin. The background was
taken from three source-free regions in the image around the cluster
core.  For the pulse profile of pulsar A and variability analysis for
pulsar H, the photon arrival times were reduced to the solar system
barycenter, using the CIAO tool {\tt axbary} assuming the DE405 JPL
solar system ephemeris.

\begin{deluxetable*}{ccccccc}
  \tabletypesize{\footnotesize} \tablecolumns{6} \tablewidth{0pc}
  \tablecaption{Net Counts and Luminosities for the pulsars in M28}
  \tablehead{ \colhead{ } &
    \colhead{Soft\tablenotemark{a}} & \colhead{Hard\tablenotemark{a}} & \colhead{$L_X$\tablenotemark{b}} & \colhead{$r_{95\%}$\tablenotemark{c}} & \colhead{$|P_{R}-P_{X}|$\tablenotemark{d}} & \colhead{Detection} \\
    \colhead{PSR J1824--2452} & \colhead{(0.3--1.5 keV)} &
    \colhead{(1.5--6 keV)} & \colhead{(0.5--8 keV)} &
    \colhead{($\arcsec$)} & \colhead{($\arcsec$)} & \colhead{} }
  \startdata
  A  & 2106 & 3800  &  1147.0 & 0.28 & 0.0  & Yes \\
  B  & $\lesssim$11   & $\lesssim$10 & $\lesssim$2 & - & - & No \\
  C  & 11   & 6     & 1.7  & 0.39 & 0.21 & Yes \\
  D  & 8    & 2     & 1.0  & 0.44 & 0.99 & Yes \\
  E  & 8    & 8     & 2.0  & 0.40 & 0.49 & ? \\
  F &  13    & 10     &  1.2  & 0.50 & 0.54 & Yes \\
  G\tablenotemark{f} & 27 & 6 & $<$10.8 & 0.33 & 1.19? & ? \\
  H & 23 &  49  & 16.9 & 0.33 & 0.10 & Yes \\
  I\tablenotemark{e} & 0 & 0  &  $\lesssim$1 & - & - & No \\
  J & 37 & 21  & 1.3  & 0.34 & 0.20 & Yes \\
  K & 26 & 20  & 5.4  & 0.35 & 0.03 & Yes \\
  L\tablenotemark{f} & 350  & 536 & $<$189.2 & 0.30 & 0.43 & No \\
 \enddata
 \tablenotetext{a}{Background-subtracted net counts extracted from
   regions enclosing $\sim$90\% of the total energy at 1.5 keV from
   the nearest X-ray source to each pulsar. For MSPs I and B, for
   which no X-ray counterpart is detected, the source counts were
   extracted from a circle of radius 1.7'' centered on the radio
   timing position.}

 \tablenotetext{b}{Estimated unabsorbed X-ray luminosity in units of $10^{30}$
   ergs s$^{-1}$ in the 0.5--8 keV band based on Cash statistic (Cash
   1979) fits to the spectra with a power-law model for
   $N_{\rm H}=2.4\times10^{21}$ cm$^{-2}$, and $D=5.5$ kpc. For MSPs
   A, G, H, and L the values were derived from spectral fitting using
   the $\chi^2$ statistic with $N_{\rm H}$ as a free parameter (see Table 3).}

 \tablenotetext{c}{Positional uncertainty (95\% confidence error
   circle radius) in arcseconds of the putative X-ray counterpart to
   each radio MSP, determined using Equation 5 from Hong et
   al.~(2005). For pulsars B and I there are no plausible X-ray
   counterparts in the vicinity.}

 \tablenotetext{d}{Angular separation in arcseconds between the radio
   timing position of each MSP (from S.~B\'egin, et al., in
   preparation) and the centroid of the nearest X-ray source. For pulsar A,
   the difference is zero as this pulsar was used to align the
   radio and X-ray astrometric frames.}

\tablenotetext{e}{Emission strongly contaminated by neighboring
   bright source.}  

\tablenotetext{f}{Possibly an unresolved blend of
   two or more unrelated sources.}
\end{deluxetable*}

\section{Imaging Analysis and Source Detection}

Figure 1 shows the co-added image of all ACIS-S exposures (237.1 ks in
total) of the core of M28 in the 0.5--2 keV band (where most of the
emission from MSPs is expected), with 1$\arcsec$ circles centered on
the 12 current best radio pulsar timing positions (S.~B\'egin, et
al.~in preparation). It is apparent that the unresolved diffuse
emission in the shallower ACIS-S observations reported by
\citet{Beck03} is mostly resolved into multiple faint sources, with
some apparently coinciding with the known radio pulsars in M28.  To
formally identify the X-ray counterparts to the pulsars, we used the
CIAO tool {\tt wavdetect}, which correlates the image with Mexican-hat
wavelets of various scales to identify sources. We also used the {\tt
  PWDetect}\footnote{See
  \url{http://www.astropa.unipa.it/progetti\_ricerca/PWDetect/} for
  details.}  script \citep{Dam97a,Dam97b}, which is generally more
effective at identifying faint sources near much brighter sources.
Subsequently, we employed the IDL script {\tt
  acis\_extract}\footnote{Available at
  \url{http://www.astro.psu.edu/xray/docs/TARA/ae\_users\_guide.html}}
\citep{Broos2010} to confirm the validity of the source detections
reported by {\tt wavdetect} and {\tt PWDetect} and refine the source
positions.

For PSR B1821--24 (henceforth MSP A, M28 A, or pulsar A) we find an
offset between the radio pulsar position from \citet{Ray08} and the
known X-ray counterpart of $\Delta\alpha=-0.15\arcsec$ and
$\Delta\delta=-0.26\arcsec$ in right ascension and declination,
respectively. Thus we apply this correction to the astrometric frame
of the X-ray dataset, which results in significantly better matches
between most of the radio MSPs and X-ray sources. The positional
uncertainties of the detected X-ray sources were computed using the
empirical relation from \citet{Hong05}, which yields the $95$\% error
radius, $r_{95\%}$.  We note that the uncertainties in the radio
timing positions for the M28 pulsars are generally
$\lesssim$0.1$\arcsec$.  Therefore, the dominant source of uncertainty
in the X-ray counterpart matches is the error in the X-ray source
position, with the exception of pulsars D and I. Pulsar D is
significantly younger (based on its characteristic age
$\tau\equiv\dot{P}/2P\approx26$ Myr) then the other pulsars in M28 and
has a comparatively long spin period of $P=79$ ms. As such, it
exhibits more timing noise, resulting in an uncertainty in its radio
position, especially in declination, comparable to $r_{95\%}$
($\sim$$0.5"-1"$). The error ellipse shown in Figure 1 was obtained by
doubling the formal uncertainties in the pulsar position obtained with
TEMPO. In the case of MSP I, the presence of random and irregular
radio eclipses, apparent long-term variations in the orbital motion,
and infrequent detections of the pulsar pose major difficulties in
deriving a unique phase-connected timing solution and thus a precise
radio timing position (see S.~B\'egin et al. in preparation, for
details). Thus, the error ellipse shown in Figure 1, taken to be five
times the formal position uncertainty derived from a TEMPO fit with a
$\chi_{\nu}^2$ over 7, represents the current best estimate of the
location of pulsar I, which may be subject to significant change.  We
note that the best timing fit of the eclipsing binary MSP H also has a
large $\chi_{\nu}^2$, implying that the uncertainties in the position
reported by TEMPO are not realistic estimates of the true
errors. Fortunately, the recent identification of its optical
counterpart by \citet{Pal10} provides a much more reliable
localization of M28H.

As evident from Figure 1 and Table 2, MSPs A, C, H, J, and K are found
well within the 95\% error circles of X-ray sources, with differences
between the X-ray and radio positions of $\lesssim$0.24$\arcsec$. Thus
we deem the X-ray identifications of these pulsars as secure.  For
pulsar D, a single X-ray source falls within the radio timing error
ellipse. As the source is found away from the crowded cluster core, it
is most likely the X-ray counterpart to the pulsar.  Pulsar F is
located just 0.04$\arcsec$ outside of $r_{95\%}$ of the sole X-ray
source in the vicinity, suggesting that the two are related. It is
likely that the small positional discrepancy is due to a slight
rotational offset between the X-ray and radio astrometric
frames. However, as pulsar A is the only bright X-ray source with a
radio counterpart in this field, it is not possible to unambiguously
correct for this effect.  Pulsar E is found 0.12$\arcsec$ outside of
the 95\% error circle of the nearest X-ray source. As this MSP is
located in the crowded cluster core, this association is less certain.
Although faint emission is coincident with the position of MSP B, the
source detection algorithms fail to identify an X-ray counterpart to
this pulsar. The detection of MSP I is rendered difficult by its
position between MSP K and one of the brightest sources in the cluster
\citep[23 in][]{Beck03}. The centroid of the X-ray counterpart to
pulsar K is nearly coincident with its radio position
($|P_{R}-P_{X}|=$0.03$\arcsec$), implying that pulsar I is either
significantly fainter than K or is not at the location derived from
radio timing.  MSP L is nearly coincident with a moderately luminous
X-ray source \citep[22 in][]{Beck03}. As discussed in \S 6, however,
based on the properties of this pulsar as inferred from radio
observations and the temporal behavior of the X-ray emission from
source 22, the two objects are most likely not physically associated.
For MSP G, three X-ray sources are found $\sim$1$\arcsec$ away (see
Figure 1), making the X-ray-radio association ambiguous as the X-ray
photons coincident with its radio position may be due to collective
emission from the three nearby X-ray sources. It is also difficult to
reliably constrain the intrinsic pulsar flux as there is no
appropriate region from which to extract a representative
background. Consequently, in order to place a crude upper limit on the
flux, we extracted source counts within 1'' of the radio position.

Given the high X-ray source number density in the cluster core, it is
important to investigate the possibility that X-ray sources coincide
with most of the radio pulsar positions purely by chance (with the
exception of M28 A, for which the radio--X-ray association is well
established). We have applied systematic offsets in the range
$0.5-5\arcsec$ in all directions of the radio pulsar positions with
respect to the X-ray dataset. We find that any offset consistently
results in two or less source matches. The same holds true for all
angles of rotation of the radio pulsar positions about the cluster
center or about the position of pulsar A. This finding suggests that
the majority of the radio--X-ray source matches are indeed genuine and
is consistent with pulsars L and/or G being possible false matches.

Table 2 summarizes the background-subtracted X-ray net source counts
extracted from polygonal regions that enclose $\sim$90\% of the total
energy at 1.5 keV for the putative X-ray counterparts of the pulsars.
For MSPs I and B, for which no X-ray counterparts are detected, we
extracted source counts from circles of radius 1.7'' (enclosing 95\%
of the energy at 1.5 keV) centered on the radio timing positions. The
background event contribution was determined from 1.7'' circles that
are the same radial distance from source 23 \citep[][]{Beck03} as the
MSPs. In both cases the net source counts are consistent with zero.

For the X-ray-faint pulsars (C, D, E, F, J, and K), we estimate the
unabsorbed X-ray luminosity by fitting a powerlaw spectrum with
$N_{H}=2.4\times10^{21}$ cm$^{-2}$ using the Cash statistic
\citep{Cash79} in {\tt XSPEC}.  This particular spectral model was
chosen because it reproduces the rough spectral shape of a
two-temperature thermal model observed in nearby MSPs
\citep{Beck02,Zavlin06,Bog09} quite well for crude photon statistics,
with spectral photon index in the range $2-3.5$.  In the case of MSPs
A, G, H, and L the luminosities were derived from the spectral fits
discussed in \S 4 and 5, although for G and L these values most
probably do not correspond to the intrinsic pulsar values as a
consequence of source confusion/blending.

We note that in the combined \textit{Chandra} HRC-S data (totaling 90
ks), of the 12 pulsars only pulsar A is detected, owing to the
$\sim$3 times lower sensitivity at 0.5-3 keV and the higher
background of HRC compared to ACIS.

\section{PSR B1821--24}

%
%
\begin{figure}[!t]
\begin{center}
\includegraphics[angle=270,width=0.46\textwidth]{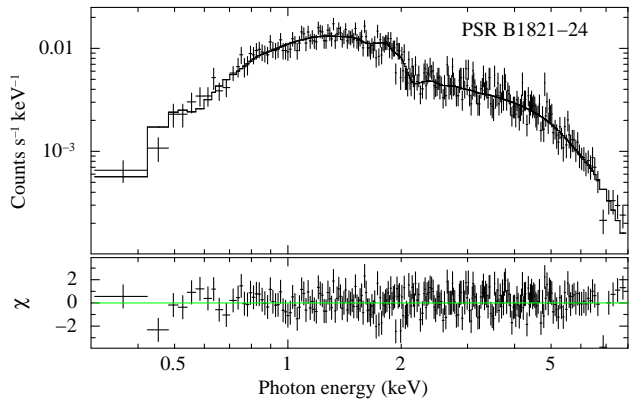}
\end{center}
\caption{Phase-integrated \textit{Chandra} ACIS-S X-ray spectrum of
  PSR B1821--24 fitted with an absorbed power-law spectrum. The lower
  panel shows the best fit residuals. See text and Table 3 for best
  fit parameters.}
\end{figure}

PSR B1821--24 is the first MSP to be discovered in a globular cluster
\citep{Lyne87}. This solitary pulsar with $P=3.05$ ms, $\dot{P}\equiv
{\rm d}P/{\rm d}t=1.61\times10^{-18}$ s s$^{-1}$, is $\sim$100 times
younger than most ``recycled'' MSPs, based on its characteristic age
($\tau_c=P/2{\dot P}=30$ Myr). It is also the most energetic
MSP known, with spin-down luminosity $\dot{E}=2.2\times10^{36}$ ergs
s$^{-1}$ and X-ray luminosity $L_X\approx1.3\times10^{33}$ ergs
s$^{-1}$ \citep[phase integrated for 0.5-8 keV, see][]{Beck03}, 2--3
orders of magnitude greater than typical MSPs. Only two other X-ray
detected MSPs, PSRs B1937+21 and J0218+4232 (both field MSPs), have
comparable values of $L_X$ \citep{Cus03,Webb04}.

Because of the bright nature of M28 A, $\sim$4\% of the X-ray photons
from this source in the ACIS-S observations are expected to suffer
from pileup (based on PIMMS\footnote{See
  \url{http://cxc.harvard.edu/toolkit/pimms.jsp.}}).  Pileup occurs
when two or more photons arrive within the same detector pixel within
a single ACIS frame integration time and are registered as a single
event. The detected energy of such an event is approximately equal to
the sum of the energies of the individual photon events of which it is
comprised. The result is a distortion of the intrinsic source spectrum
\citep{Davis01}. To ascertain the severity of this effect we have
included the {\tt pileup} model provided in XSPEC in our spectral
fits. We find that pileup has negligible impact on the spectrum of PSR
B1821--24, as the best fit values of the model parameters with and
without pileup are well within 1$\sigma$ of one another. Based on
this, in what follows, we ignore the effect of pileup.

As expected, the phase-integrated X-ray spectrum of PSR B1821--24 is
very well described by a pure absorbed power-law (Figure 2). The best
fit parameters are $\Gamma=1.226^{+0.025}_{-0.028}$, $N_{\rm
  H}=(2.55^{+0.13}_{-0.12})\times10^{21}$ cm$^{-2}$ with
$\chi^2_{\nu}=0.93$ for 253 degrees of freedom. The uncertainties
quoted are 1$\sigma$ for one interesting parameter. The unabsorbed
energy flux is $F_X=(3.80^{+0.13}_{-0.09})\times10^{-13}$ ergs
s$^{-1}$ (0.3--8 keV), corresponding to a luminosity of
$L_X=1.4\times10^{33}\Theta(D/5.5~{\rm kpc})^2$ ergs s$^{-1}$, where
$\Theta$ is the (unknown) portion of the sky covered by the X-ray
emission beams.

By analogy with most MSPs, M28 A may have heated magnetic polar
caps. As a result, low-level thermal emission could be present in the
predominantly non-thermal spectrum of PSR B1821--24. Introducing
either a blackbody or hydrogen atmosphere model \citep[NSATMOS
  from][]{McC04} with temperatures in the range $(0.5-3)\times10^6$ K
yields acceptable fits, although given the high-quality of the pure
power-law fit, the addition of the thermal component is not warranted
by the data (as confirmed by an F-test).  This allows us to set a firm
upper bound on any thermal component of $\sim$1\% of the total energy
flux in the 0.3--8 keV range. The implied thermal luminosity of
$\lesssim$$10^{31}$ ergs s$^{-1}$ is comparable to most nearby MSPs
\citep{Zavlin06,Bog09} and those in 47 Tuc \citep{Bog06}.  On the
other hand, in some theoretical pulsar models \citep[see,
  e.g.,][]{Hard02}, the energetics of the magnetosphere of PSR
B1821--24 may not be conducive to the production of a substantial
return current of particles to heat the polar cap surface so this
pulsar may in fact have much cooler polar caps.

\citet{Beck03} have reported intriguing but marginal evidence (at 98\%
confidence) for a broad emission feature centered at 3.3 keV, which
they interpreted as cyclotron emission from the magnetosphere above
the pulsar polar cap if the magnetic field strongly deviates from a
centered dipole.  We have reprocessed the archival ACIS-S data using
the latest Chandra X-ray Center calibration products and find no
evidence of this feature. As seen from Figure 2, the broad spectral
feature is absent in the total ACIS-S spectrum as well. Therefore, we
attribute the deviation from a pure power-law of the PSR B1821--24
spectrum found by \citet{Beck03} to an unmodelled instrumental feature
that has since been incorporated into the telescope calibration model.

\subsection{The Unpulsed X-ray Emission from PSR B1821--24}

The bulk of X-rays from PSR B1821--24 ($\sim$85\%) are emitted in two
very narrow pulses \citep{Rut04}, indicative of highly beamed,
non-thermal radiation from the pulsar magnetosphere, in stark contrast
with the broad, soft thermal pulsations of most MSPs
\citep{Zavlin06,Bog06,Cam07,Bog09}.  Surprisingly, \citet{Rut04} have
found an appreciable unpulsed component, contributing $\sim$15\% of
total photon flux.

Using the two HRC-S timing observations of PSR B1821--24 we can place
constraints on the origin of the puzzling unpulsed emission. To this
end, we combined the pulse profiles from the two HRC datasets as
follows.  We first folded the barycentered arrival times of the events
extracted from a 1$\arcsec$ region centered on the radio pulsar in
TEMPO\footnote{Available at \url{http://tempo.sourceforge.net}.} using
the timing ephemeris from \citet{Ray08} for each dataset separately.
Due to the $\sim$3.5-year gap between the two HRC-S observations it is
not possible to phase-connect the two datasets accurately using the
available radio timing ephemerides. Thus, to determine the relative
phase misalignment, we cross-correlated the folded pulse profiles from
both observations. The individual and combined pulse profiles are
shown in Figure 3.  The summed HRC-S pulse profile yields 706 photons,
a 1.8-fold improvement in photon statistics over the data from
\citet{Rut04}. Of these, $16\pm0.4$ photons are due to background.  As
in \citet{Rut04}, the unpulsed (DC) level in the summed profile was
determined by considering the emission $\pm0.15$ in phase away from
the peaks of the two pulsas to minimize the contribution from the
pulsed component. After subtracting the background contribution, we
find that $17.5\%\pm4\%$ of the source photons are unpulsed. Varying
the choice of phase intervals used to determine the unpulsed level
results in a spread of $\pm1$\% around the nominal value.

%
%
\begin{figure}[!t]
\begin{center}
\includegraphics[width=0.46\textwidth]{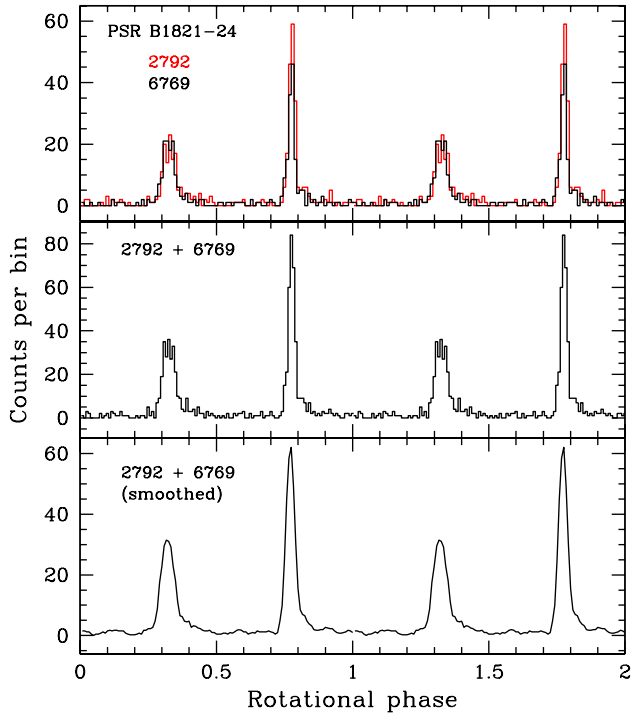}
\end{center}
\caption{\textit{Chandra} HRC-S pulse profiles of PSR B1821--24 from
  ObsIDs 2797 and 6769 (\textit{top panel}) and the summed pulse
  profile (\textit{middle panel}). The \textit{bottom} panel shows the
  summed profile smoothed using a 4-point moving average. Note the
  asymmetry of the wings of the two pulses. The choice of phase zero
  is arbitrary. Two cycles are shown for clarity.}
\end{figure}

Although the HRC has poor intrinsic energy resolution, in principle,
it is still possible to determine any gross difference between the
spectra of the pulsed and unpulsed emission using hardness
ratios\footnote{See
  \url{http://cxc.harvard.edu/proposer/POG/html/chap7.html} for
  details.}. The hardness ratios, however, do not show any substantial
differences in the spectra of the pulsed and unpulsed portions,
implying that the unpulsed emission is not drastically softer or
harder than the pulsed emission.  Nonetheless, the high-quality
phase-integrated ACIS-S continuum provides stringent constraints on
the spectral properties of the unpulsed emission, in particular, if it
is of a thermal nature or not. The best-fit absorbed power-law model
derived from the ACIS-S spectrum, convolved with the HRC-S effective
area curve, reveals that the bulk of photons from PSR B1821--24 in the
HRC-S datasets are from the 0.4--2 keV range. We fitted the ACIS-S
spectrum with additional thermal components to determine their photon
flux contribution in this energy range. Converting the resulting
values to the equivalent HRC-S count rates indicates that any thermal
emission can contribute no more than $\sim$6\% of the total number of
photons from M28 A in the HRC-S observations. This leaves at
least $\sim$7--15\% of the photons unaccounted for.

It is interesting to speculate on the origin of this excess unpulsed
emission.  In many young, energetic pulsars, with the Crab pulsar
being the most notable example \citep{Kir06}, the high-energy
non-thermal pulse profiles often exhibit a ``bridge'' of emission
connecting two pulses, suggesting that the two pulses originate from
the same emission cone. However, for PSR B1821--24 the count rates in
the regions between the two pulses are consistent with being
identical. The apparent lack of excess of photons in one of the
regions implies that bridge emission is not responsible for the
seemingly unpulsed photons.  It should be noted however that both
pulses exhibit tails of emission from their trailing edges (see bottom
panel of Figure 4), which could plausibly extend to cover most of the
rotational period.  Indeed, certain high-energy emission models of
pulsars, such as the two-pole caustic model, predict low-level
non-thermal magnetospheric emission spanning the entire period of
rotation \citep{Vent10}.

A plausible alternative origin for the unpulsed photons is synchrotron
emission from an unresolved wind nebula surrounding this rather
energetic pulsar. By analogy with pulsars with comparable spin-down
luminosities \citep[see][for a review]{Kar08}, the expected angular
size of a wind nebula surrounding pulsar M28 A, assuming a distance
of 5.5 kpc, would be $\sim$1$\arcsec$. Such a nebula would be
difficult to resolve, especially considering the source crowding in
the cluster core.  If the unpulsed emission arises solely
due to the wind nebula, its implied X-ray luminosity is $L_{\rm
  PWN}\sim10^{32}$ ergs s$^{-1}$, similar to what is observed for
other energetically comparable pulsars \citep[see in particular Tables
1 and 2 in][]{Kar08}.

Conceivably, the unpulsed emission could be caused by small-angle
scattering of X-rays passing through dust grains in the interstellar
medium \citep{Smith02}. The additional path length of the scattered
X-rays would induce a time delay in the photon time of arrival,
causing them to become out of phase relative to the unscattered
photons and thus form an unpulsed component. This effect could also
account for the trailing tails of the two pulses.  The location of M28
in the Galactic disk ($l=7.8^{\circ}$, $b=5.6^{\circ}$) suggests that
dust scattering is not neglegible. We estimate the optical depth to
scattering using the empirical relation $\tau_{\rm scat,
  1keV}=A_V(0.079\pm0.003)-(0.052\pm0.019)$ \citep{Pred95}, where
$A_V$ is the visual interstellar extinction. For $A_V\sim1.4$ mag
towards M28 \citep{Davidge96}, this yields $\tau\sim0.05$ implying
that $I/I_o=e^{-\tau}\sim5$\% of the X-ray photons undergo scattering.

Finally, it is possible that at least a portion of the unpulsed
component from PSR B1821-24 is due to one or more fainter, unrelated
X-ray sources within $\sim$1$\arcsec$ of PSR B1821--24. However, using
{\tt wavdetect} we find that the centroid of the unpulsed emission is
offset by only 0.07$"$ from the centroid of the total emission,
suggesting that source contamination is unlikely.

%
%
\begin{figure}[!t]
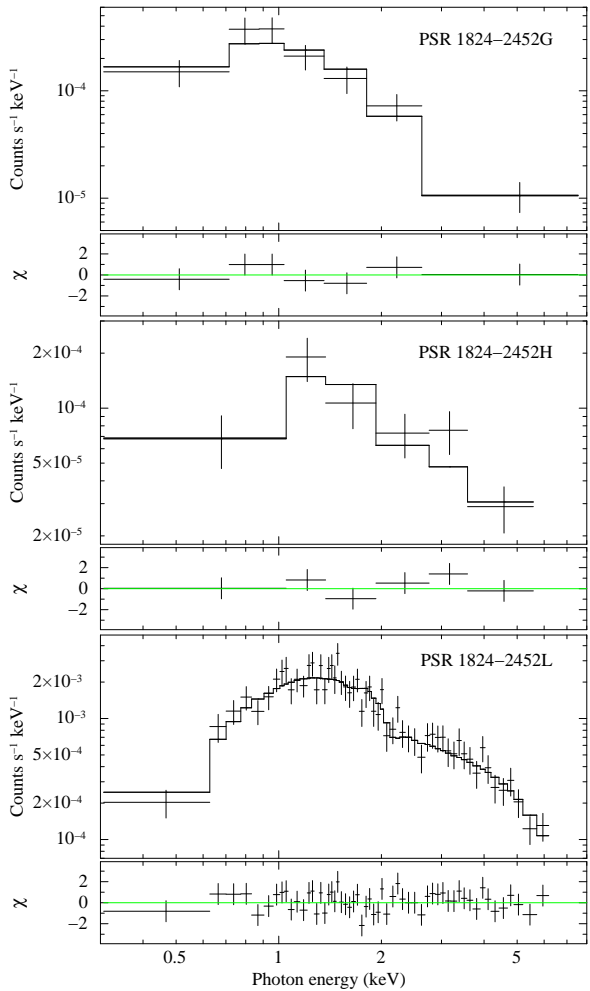

\begin{center}
\includegraphics[angle=270,width=0.43\textwidth]{f4a.ps}
\includegraphics[angle=270,width=0.43\textwidth]{f4b.ps}
\includegraphics[angle=270,width=0.43\textwidth]{f4c.ps}
\end{center}
\caption{X-ray spectra of PSR J1824-2452H and of the X-ray sources
  nearest to PSRs J1824-2452G and L, fitted with power-law spectra. The
  bottom panels show the best fit residuals. See Table 3 for best fit
  parameters.}
\end{figure}

\begin{deluxetable}{lcccc}
\tabletypesize{\small} 
\tablecolumns{5} 
\tablewidth{0pc}
\tablecaption{Spectral fits for M28 MSPs A, G, H, and L.}
\tablehead{ \colhead{MSP} & \colhead{{\bf A}} & \colhead{G\tablenotemark{a}} & \colhead{{\bf H}} & \colhead{L\tablenotemark{a}} }
\startdata
Powerlaw & & & & \\
\hline
$N_{\rm H}$\tablenotemark{b} & ${\bf 2.55}^{{\bf +0.13}}_{{\bf -0.12}}$ & $1.25^{+0.60}_{-0.78}$ & ${\bf 2.1}^{{\bf +1.5}}_{{\bf -1.1}}$ & $3.2^{+0.4}_{-0.3}$ \\
$\Gamma$  & ${\bf 1.226}^{{\bf +0.025}}_{{\bf -0.028}}$ &  $2.12^{+0.21}_{-0.25}$ & ${\bf 1.09}^{{\bf +0.28}}_{{\bf -0.29}}$ & $1.52^{+0.07}_{-0.08}$ \\
$F_X$\tablenotemark{c}  & ${\bf 380}^{{\bf +13}}_{{\bf -9}}$ & $3.6^{+0.6}_{-1.1}$ & ${\bf 4.8}^{{\bf +0.9}}_{{\bf -3.6}}$ & $55.4^{+4.9}_{-5.0}$ \\
$\chi^2_{\nu}/$d.o.f. & ${\bf 0.93/253}$ & $0.89/4$ & ${\bf 0.87/3}$ & $0.82/54$
\enddata
\tablenotetext{a}{Spectrum likely dominated by unrelated X-ray source(s).}
\tablenotetext{b}{Hydrogen column density along the line of sight in units of $10^{21}$ cm$^{-2}$.}
\tablenotetext{c}{Unabsorbed flux in units of $10^{-15}$ ergs cm$^{-2}$ s$^{-1}$ in the 0.3--8 keV band.}
\end{deluxetable}

\section{PSR J1824--2452H}

Apart from PSR B1821--24, the binary PSR J1824--2452H is the only
other unambiguously detected MSP with sufficient counts for a spectral
fitting analysis.  PSR J1824--2458H is in a 10.4-hour circular orbit
around what appears to be a non-degenerate star \citep{Pal10} and
exhibits highly variable and irregular radio eclipses (S.~B\'egin et
al. in preparation). This atypical MSP binary system is thus likely
the result of a past dynamical binary exchange encounter in the dense
cluster core \citep{Cam05}, like PSR J1740--5340 in NGC 6397
\citep{DAm01,Grind02,Bog10,Huang10} and PSR J0024-7204W in 47 Tuc
\citep{Cam00,Edm02,Bog05}. Alternatively, it may have just emerged
from the low-mass X-ray binary phase with the MSP re-activating as a
rotation-powered pulsar, as is also possibly the case for PSR
J1023+0038 \citep[see][]{Arch09}.  The X-ray properties of PSR
J1824--2452H are thus of particular interest as these systems show
unusual X-ray properties \citep{Stap03,Bog05,Bog10,Arch10}.  We fitted
the spectrum of this MSP with a power-law model (middle panel of
Figure 4), yielding $N_{\rm H} =(2.1^{+1.5}_{-1.1})\times10^{21}$
cm$^{-2}$, $\Gamma=1.09^{+0.28}_{-0.29}$, and unabsorbed flux of
$F_X=(4.8^{+0.9}_{-3.6})\times10^{-15}$ (0.3--8 keV), with
$\chi_{\nu}^{2}=0.87$ for 5 degrees of freedom. The nominal best fit
flux translates to a luminosity of $L_X=1.7\times10^{31}$ ergs
s$^{-1}$ for a 5.5 kpc distance.  Despite the limited photon
statistics the spectrum is poorly reproduced by a single-temperature
thermal model. Applying models with two or more components offers no
meaningful constraints on any model parameters.

%
%
\begin{figure}[t]
\begin{center}
\includegraphics[width=0.48\textwidth]{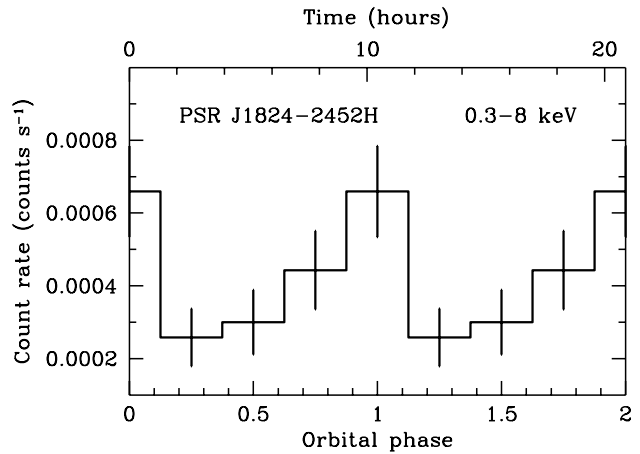}
\end{center}
\caption{X-ray lightcurve of PSR J1824--2452H folded at the 10.4-hour
  binary period of the pulsar. The orbital phase is defined based on
  the radio pulsar timing convention in which superior conjunction
  occurs at $\phi=0.25$. Two orbital cycles are shown for clarity.}
\end{figure}

For this binary MSP we can also investigate any large amplitude X-ray
variability as a function of orbital phase, which could potentially
provide clues regarding the origin of the X-ray emission.  We have
used the radio timing ephemeris of PSR J1824--2452H (S.~B\'egin et
al., in preparation) to fold the X-ray dataset at the binary
period. The resulting lightcurve (Figure 5) shows compelling evidence
for variability. For four and five phase bins over the binary orbit,
the variability is significant at a 4$\sigma$ and 2.7$\sigma$ level,
respectively, based on a $\chi^2$ test. A Kuiper test \citep{Pal04} on
the unbinned lightcurve, weighted to account for the non-uniform
exposure across the orbit, indicates a $5.8\times10^{-4}$
(3.2$\sigma$) probability that photons being drawn from a constant
distribution would exhibit this level of non-uniformity. It is
interesting to note that the minimum around $\phi=0.25$ in the X-ray
lightcurve is roughly coincident with the position of the radio
eclipse. Moreover, the rough shape of the lightcurve is remarkably
similar to those of the analogous binaries PSRs J0024--7204W
\citep{Bog05}, J1740--5340 \citep{Bog10}, and J1023+0038
\citep{Arch10}.  Thus, the peculiar nature of the PSR J1824--2452H
binary system, the implied X-ray luminosity, relatively hard spectrum
compared to most MSPs observed in X-rays, and the suggested
variability at the binary period point to an intrabinary shock origin
of the X-rays associated with this object. The minimum in the X-ray
light curve at $\phi\approx0.25$ can be plausibly interpreted as a
geometric occultation (either partial or total) of the shock by the
secondary star. Alternatively, it could arise due to relativistic
beaming of the X-rays from the radiating particles in the shock.  Due
to the limited photon statistics and the number of orbits covered
($\sim$6) it is difficult to establish whether the flux modulation is
truly periodic and stable over many orbits. As seen in PSR
J0024--7204W \citep{Bog05,Cam07}, the orbital variation of the X-ray
flux may actually evolve on timescales spanning from months to years.

%

\section{Other MSPs}

The $\sim$10 times larger absorbing column towards M28 compared to 47
Tuc and $\sim$1.2 times greater distance make it more difficult to
study soft X-ray sources such as thermal MSPs \citep{Bog06} in a
comparable exposure. As a result, most of the detected pulsars in M28
do not have adequate photon statistics for detailed spectral analyses.
Nonetheless, we can still constrain the origin of the observed X-rays
by considering the summed spectrum of pulsars B, C, D, E, F, J, and K
(shown in Figure 6).  Indeed, the combined X-ray continuum of these
seven pulsars is poorly described by a single-temperature thermal
spectrum, while a pure power-law model requires $N_{\rm H} > 4
\times10^{21}$ cm $^{-2}$ for an acceptable fit, substantially higher
than the value suggested by the fit to the spectrum of PSR B1821--24.
A composite blackbody plus power-law model for
$N_{\rm}=2.4\times10^{21}$ cm$^{-2}$ gives an acceptable fit with
$kT=0.26\pm0.03$ keV, $R_{\rm eff}=0.19^{+0.11}_{-0.13}$ km,
$\Gamma=0.8^{+0.9}_{-1.5}$, and $\chi_{\nu}^2=0.86$ for 10 degrees of
freedom. Note that a power-law component does not necessarily imply
non-thermal emission since the superposition of several
multi-temperature thermal MSP spectra can produce this spectral shape
for limited photon statistics \citep{Zavlin06,Bog09}.  The derived
unabsorbed flux of $7.5\times10^{-15}$ ergs cm$^{-2}$ s$^{-1}$ (0.3--8
keV) implies a mean X-ray luminosity of $3.9\times10^{30}$ ergs
s$^{-1}$ for the seven pulsars or $4.5\times10^{30}$ ergs s$^{-1}$ if
we exclude pulsar B. For comparison, the mean unabsorbed X-ray
luminosity of the 11 thermal MSPs in 47 Tuc with unconfused and
uncontaminated spectra is $\sim3.2\times10^{30}$ ergs s$^{-1}$
\citep{Bog06}. This suggests that the MSP populations of the two
clusters are quite similar in terms of X-ray properties.

The non-detection of pulsars B and I implies that these pulsars may be
below the sensitivity of the total ACIS-S exposure, with $L_X\lesssim
1\times10^{30}$ ergs s$^{-1}$ (0.3--8 keV), perhaps similar to the
unusually sub-luminous PSR B1257+12 with $L_X\sim2.5\times10^{29}$
ergs s$^{-1}$ in the same energy band \citep{Pav07}. The low
luminosity of the non-recycled pulsar D is similar to those of
comparably old nearby pulsars \citep[see][]{Kar06}.

For pulsars G and L, there are sufficient counts within 1$\arcsec$ to
conduct spectral fitting and search for variability, although as noted
in \S 3, the association of this emission with the pulsars is
questionable. For both sources, the spectra are well described by an
absorbed power-law model (see Table 3 and Figure 4). In the case of
pulsar G, there is no statistically significant variability. As for
PSR J1824--2452L, the X-rays extracted from within 1$\arcsec$ of the
radio timing position exhibit variability on time-scales of order
hours. In addition, there is a substantial increase (a factor of
$\sim$3) in the count rate from the three 2002 ACIS-S observations to
the two observations in 2008. In principle, such large-amplitude
variability from an MSP binary could arise due to interaction of the
pulsar wind with gas from the companion star. However, this MSP is not
unusual and exhibits no radio eclipses, which are associated with
intrabinary shocks in all known cases. Therefore, the variable X-rays
are most probably from source 22 from \citet{Beck03}, which
is not related to the pulsar.

\section{CONCLUSION}

\begin{figure}[!t]
\begin{center}
\includegraphics[angle=270,width=0.43\textwidth]{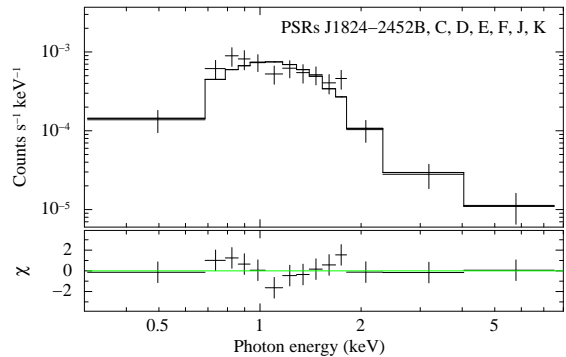}
\end{center}
\caption{Summed X-ray spectrum of PSRs J1824--2452B, C, D, E, F, J,
  and K fitted with a blackbody plus powerlaw model. The bottom panels
  show the best fit residuals. See text for best fit parameters.}
\end{figure}

We have presented deep \textit{Chandra} observations of the sample of
known pulsars in the globular cluster M28.  Seven of the 12 known
pulsars in the cluster are securely detected (A, C, D, F, H, J, and K)
two are possibly detected (E and G), while three are not detected (B,
L, and I).  Most of the MSPs firmly or possibly detected appear to be
relatively faint and soft sources most likely due to thermal polar cap
emission.  In this sense, they appear to be akin to the majority of
``recycled'' MSPs in 47 Tuc \citep{Bog06} and the field of the Galaxy
\citep{Zavlin06,Bog09}, with soft spectra and
$L_X\sim4\times10^{30-31}$ ergs s$^{-1}$ (0.3--8 keV).  This provides
further evidence that in typical MSPs, both in globular clusters and
the field of the Galaxy, the X-ray emission is primarily of a thermal
nature. As shown by \citet{Pavlov97}, \citet{Zavlin98}, \citet{Bog07},
and \citet{Bog08}, realistic modelling of the X-ray spectra and
pulsations of these objects may potentially allow stringent
constraints on the neutron star equation of state through measurement
of the mass-to-radius ratio of these stars.As such, MSPs warrant
further investigation in the X-ray band.

The most detailed X-ray spectrum to date of the energetic PSR
B1821--24 is found to be well described by a pure power-law with
$\Gamma=1.2$, with no requirements for additional components. We set a
limit on any thermal X-ray emission from the MSP of
$\lesssim$$10^{31}$ ergs s$^{-1}$ for 0.3--8 keV. The greatly improved
ACIS-S spectrum of this pulsar shows no evidence for the previously
reported spectral feature at $\sim$3.3 keV.  A re-examination of the
X-ray pulse profile of PSR B1821--24 reveals that the observed
unpulsed emission is unlikely to be of thermal origin and is due to
either low-level non-thermal magnetospheric emission, unresolved
nebular emission, and/or dust scattering of the pulsed X-rays.

PSR J1824--2452H appears to share common X-ray characteristics
(namely, variable and hard non-thermal emission with $L_X\sim10^{31}$
ergs s$^{-1}$) with PSRs J0024--7204W, J1023+0039, and J1740--5340.
Presumably, in these peculiar binary systems the X-rays originate from
an intrabinary (synchrotron) X-ray emitting shock formed by
interaction of the MSP wind and matter from the secondary star
\citep{Arons93}.  Furthermore, this class of objects exhibits
remarkably similar X-ray and optical properties to the accreting X-ray
MSPs in quiescence \citep{Camp04, Hein07,Hein09} and thus can
potentially offer vital clues into the little understood activation
mechanism for rotation-powered MSPs.

\acknowledgements We would like to thank Z.~Medin, A.~Harding, and
A.~Archibald for useful discussions and the referee F.~Camilo for
comments. The work presented was funded in part by NASA
\textit{Chandra} grant GO8-9072X awarded through the Harvard College
Observatory. S.B. is supported in part by a CIFAR Junior
Fellowship. C.O.H. and I.H.S. are supported by NSERC.  This research
has made use of the NASA Astrophysics Data System (ADS) and software
provided by the Chandra X-ray Center (CXC) in the application package
CIAO.

Facilities: \textit{CXO}


\begin{thebibliography}{}


\bibitem[Abdo et al.(2009)]{Abdo09} Abdo, A.~A., et al.~2009, Science, 325, 848

\bibitem[Archibald et al.(2009)]{Arch09} Archibald, A.~M., et al.~2009, Science, 324, 1411

\bibitem[Archibald et al.(2010)]{Arch10} Archibald, A.~M., Kaspi, V.~M., Bogdanov, S., Hessels, J.~W.~T., Stairs, I.~H., Ransom, S.~M., McLaughlin, M.~A., \& Lorimer, D.~2010, ApJ, 722, 88

\bibitem[Arons \& Tavani(1993)]{Arons93} Arons, J. \& Tavani M. 1993, \apj, 403, 249

\bibitem[Bassa et al.(2004)]{Bass04a} Bassa, C.~G., Pooley, D., Homer, L., Verbunt, F., Gaensler, B.~M., Lewin, W. H. G., Anderson, S.~F., Margon, B., Kaspi, V.~M., \& van der Klis, M. 2004, \apj, 609, 755

\bibitem[Becker \& Aschenbach(2002)]{Beck02} Becker, W. \& Aschenbach,
B. 2002, Proceedings of the 270. WE-Heraeus Seminar on Neutron Stars,
Pulsars, and Supernova Remnants, Eds. W. Becker, H. Lech,
J. Tr\"umper, p.~64

\bibitem[Becker et al.(2003)]{Beck03} Becker, W., Swartz, D. A., Pavlov, G. G., Elsner, R. F., Grindlay, J. E., Mignani, R., Tennant, A. F., Backer, D., Pulone, L., Testa, V., \& Weisskopf, M. C. 2003, \apj,  594, 798

\bibitem[B\'egin(2006)]{Beg06} B\'egin, S., Thesis submitted to the
  Faculty of Physics, University of British Columbia, 2006

\bibitem[Bogdanov et al.(2005)]{Bog05} Bogdanov, S., Grindlay, J. E., \& van den Berg, M. 2005, \apj, 630, 1029 

\bibitem[Bogdanov et al.(2006)]{Bog06} Bogdanov, S., Grindlay,
J.~E., Heinke, C.~O., Camilo, F., Freire, P.~C.~C., \& Becker, W. 2006,
\apj, 646, 1104

\bibitem[Bogdanov et al.(2007)]{Bog07} Bogdanov, S.,
Rybicki, G. B., \& Grindlay, J. E.  2007, \apj, 670, 668

\bibitem[Bogdanov et al.(2008)]{Bog08} Bogdanov, S., Grindlay, J. E.,
\& Rybicki, G. B.  2008, \apj, 689, 407

\bibitem[Bogdanov \& Grindlay(2009)]{Bog09} Bogdanov, S. \& Grindlay, J.~E. 2009, \apj, 703, 1557

\bibitem[Bogdanov et al.(2010)]{Bog10} Bogdanov, S., van den Berg, M., Heinke, C.~O., Cohn, H.~N., Lugger, P.~M., \& Grindlay, J.~E. 2010, \apj, 709, 241

\bibitem[Broos et al.(2010)]{Broos2010} Broos, P.~S., Townsley, L.~K., Feigelson, E.~D., Getman, K.~V., Bauer, F.~E., \& Garmire, G.~P.\ 2010, arXiv: 1003.2397

\bibitem[Cameron et al.(2007)]{Cam07} Cameron, P.~B., Rutledge, R.~E.,
  Camilo, F ., Bildsten, L., Ransom, S.~M., \& Kulkarni, S.~R. 2007,
  \apj, 660, 587

\bibitem[Camilo et al.(2000)]{Cam00} Camilo, F., Lorimer, D. R., Freire, P., Lyne, A. G., \& Manchester, R. N.  2000, \apj, 535, 975

\bibitem[Camilo \& Rasio(2005)]{Cam05} Camilo. F. \& Rasio, F.~A. 2005, ASP Conf. Ser. Vol. 328: Binary Radio Pulsars, eds. F.~A. Rasio \& I.~H. Stairs (San Francisco: ASP), p.~147

\bibitem[Campana et al.(2004)]{Camp04} Campana, S., D'Avanzo. P., Casares, J., Covino, S., Israel, G.~L., Marconi, G., Hynes, R., Charles, P., Stella, L. 2004, \apj, 614, L49

\bibitem[Cash(1979)]{Cash79} Cash, W. 1979, \apj, 228, 939

\bibitem[Cusumano et al.(2003)]{Cus03} Cusumano, G., et al. 2003, A\&A, 410, L9

\bibitem[D'Amico et al.(2001)]{DAm01} D'Amico, N., Lyne, A. G., Manchester, R. N., Possenti, A., \& Camilo, F. 2001, \apj, 548, L171

\bibitem[D'Amico et al.(2002)]{DAmico02} D'Amico, N., Possenti, A., Fici, L., Manchester, R. N., Lyne, A. G., Camilo, F., \& Sarkissian, J. 2002, \apj, 570, L89

\bibitem[Damiani et al.(1997a)]{Dam97a} Damiani, F., Maggio, A., Micela, G., \& Sciortino, S. 1997a, \apj, 483, 350

\bibitem[Damiani et al.(1997b)]{Dam97b} Damiani, F., Maggio, A., Micela, G., \& Sciortino, S. 1997b, \apj, 483, 370

\bibitem[Davidge(1996)]{Davidge96} Davidge, T.~J. 1996, \apj, 468, 641 

\bibitem[Davis(2001)]{Davis01} Davis, J.~E. 2001, \apj, 562, 575

\bibitem[Edmonds et al.(2002)]{Edm02} Edmonds, P.~D., Gilliland, R.~L., Camilo, F., Heinke, C.~O., \& Grindlay, J.~E. 2002, \apj, 579, 741

\bibitem[Elsner et al.(2008)]{Els08} Elsner, R.~F., et al.~2008, \apj, 687, 1019

\bibitem[Freire et al.(2003)]{Freire03} Freire, P.~C., Camilo, F., Kramer, M., Lorimer, D.~R., Lyne, A.~G., Manchester, R.~N., \& D' Amico, N.~2003, \mnras, 340, 1359

\bibitem[Freire et al.(2008a)]{Freire08a} Freire, P.~C.~C., Ransom,
  S.~M., B\'egin, S., Stairs, I.~H., Hessels, J.~W.~T., Frey, L.~H.,
  Camilo, F. 2008, \apj, 675, 670
	
\bibitem[Freire et al.(2008b)]{Freire08b} Freire, P.~C.~C., Wolszczan, A., van den Berg, M., Hessels, J.~W.~T. 2008, \apj, 679, 1433

\bibitem[Fruchter et al.(1988)]{Fru88} Fruchter, A.~S., Stinebring, D.~R., \& Taylor, J.~H. 1988, Nature, 333, 237

\bibitem[Grindlay et al.(2002)]{Grind02} Grindlay, J. E., Camilo, F., Heinke, C. O., Edmonds, P.~D., Cohn, H., \& Lugger, P. 2002, \apj, 581, 470

\bibitem[Harding \& Muslimov(2002)]{Hard02} Harding, A.~K. \&
Muslimov, A.~G. 2002, \apj, 568, 862

\bibitem[Harding et al.(2005)]{Hard05} Harding, A.~K., Usov, V.~V.,
  Muslimov, A.~G. 2005, \apj, 622, 531

\bibitem[Harris(1996)]{Harris96} Harris, W.~E. 1996, AJ, 112, 1487

\bibitem[Heinke et al.(2006)]{Hein06} Heinke, C.~O., Wijnands, R., Cohn, H.~N., Lugger, P.~M., Grindlay, J.~E., Pooley, D., \& Lewin, W.~H.~G. 2006, \apj, 651, 1098

\bibitem[Heinke et al.(2007)]{Hein07} Heinke, C.~O., Jonker, P.~G., Wijnands, R., Taam, R.~E. 2007, \apj, 660, 1424

\bibitem[Heinke et al.(2009)]{Hein09} Heinke, C.~O., Jonker, P.~G., Wijnands, R., Deloye, C.~J., \& Taam, R.~E. 2009, \apj, 691, 1035

\bibitem[Hong et al.(2005)]{Hong05} Hong, J., van den Berg, M,
  Schlegel, E.~M., Grindlay, J.~E., Koenig, X., Laycock, S., \& Zhao,
  P. 2005, \apj, 635, 907

\bibitem[Huang \& Becker(2010)]{Huang10} Huang, R.~H.~H., \& Becker, W.~2010, A\&A, 510, 67

\bibitem[Kargaltsev et al.(2006)]{Kar06} Kargaltsev, O., Pavlov, G.~G., \& Garmire, G.~P. 2006, \apj, 636, 406

\bibitem[Kargaltsev \& Pavlov(2008)]{Kar08} Kargaltsev, O., \& Pavlov, G.~G. 2008, AIPC, 983, 171
	
\bibitem[Kirsch et al.(2006)]{Kir06} Kirsch, M.~G.~G., et al. 2006, A\&A, 453, 173

\bibitem[Lyne et al.(1987)]{Lyne87} Lyne, A.~G., Brinklow, A., Middleditch, J., Kulkarni, S.~R., Backer, D.~C. 1987, Nature, 328, 399

\bibitem[McClintock et al.(2004)]{McC04} McClintock, J.~E., Narayan,
  R., \& Rybicki, G.~B. 2004, \apj, 615, 402

\bibitem[Pallanca et al.(2010)]{Pal10} Pallanca, C., Dalessandro, E., Ferraro, F. R., Lanzoni, B., Rood, R. T., Possenti, A., D'Amico, N., Freire, P. C., Stairs, I., Ransom, S. M., Bégin, S~2010, \apj, 725, 1165

\bibitem[Paltani(2004)]{Pal04} Paltani, S.~2004, A\&A, 420, 789

\bibitem[Pavlov \& Zavlin(1997)]{Pavlov97} Pavlov, G.~G. \& Zavlin,
V. E. 1997, \apj, 490, L91

\bibitem[Pavlov et al.(2007)]{Pav07} Pavlov, G.~G., Kargaltsev, O., Garmire, G.~P., \& Wolszczan, A. 2007, \apj, 664, 1072

\bibitem[Pellizzoni et al.(2009)]{Pel09} Pellizzoni, A., et al.~2009, \apj, 695, L115

\bibitem[Predehl \& Schmitt(1995)]{Pred95} Predehl, P. \& Schmitt, J.~H.~M.~M. 1995, A\&A, 293, 889

\bibitem[Ransom et al.(2005)]{Ran05} Ransom, S.~M., Hessels, J.~W.~T., Stairs, I.~H., Freire, P.~C.~C., Camilo, F., Kaspi, V.~M., Kaplan, D.~L. 2005, Science, 307, 892

\bibitem[Ray et al.(2008)]{Ray08} Ray, P.~S., Wolff, M.~T., Demorest, P., Cognard, I., Backer, D.~C., Wood, K.~S. 2008, AIPC, 983, 157

\bibitem[Rutledge et al.(2004)]{Rut04} Rutledge, R. E., Fox, D. W., Kulkarni, S. R., Jacoby, B. A., Cognard, I., Backer, D. C., \& Murray, S. S. 2004, \apj, 613, 522

\bibitem[Shawl \& White(1986)]{Shawl86} Shawl, S.~J. \& White, R.~E.~1986, AJ, 91, 312

\bibitem[Smith et al.(2002)]{Smith02} Smith, R.~K., Edgar, R.~J., \&
  Shafer, R.~A. 2002, \apj, 581, 562

\bibitem[Stappers et al.(2003)]{Stap03} Stappers, B. W., Gaensler, B. M., Kaspi, V. M., van der Klis, M., \& Lewin, W. H. G. 2003, Science, 299, 1372

\bibitem[Venter et al.(2010)]{Vent10} Venter, C., Harding, A.~K., \& Guillemot, L. 2010, \apj, 707, 800

\bibitem[Webb et al.(2004)]{Webb04} Webb, N.~A., Olive, J. -F., \& Barret, D. 2004, A\&A, 417, 181

\bibitem[Zavlin \& Pavlov(1998)]{Zavlin98} Zavlin, V. E. \& Pavlov, G. G. 1998, A\&A, 329, 583

\bibitem[Zavlin(2006)]{Zavlin06} Zavlin, V. E. 2006, \apj, 638, 951

\bibitem[Zavlin(2007)]{Zavlin07} Zavlin, V.~E. 2007, Ap\&SS, 308, 297

\end{thebibliography}
\end{document}